\documentclass[aps,prd,preprint,amssymb,eqsecnum,nofootinbib,floatfix]{revtex4}
\usepackage{epsfig}
\usepackage{amsmath}
\usepackage{amsfonts}
\usepackage{bm}
\usepackage{revsymb}
\usepackage{graphicx,epsfig}
\usepackage{placeins}
\newcommand{\be}{\begin{equation}}
\newcommand{\ee}{\end{equation}}
\newcommand{\bea}{\begin{eqnarray}}
\newcommand{\eea}{\end{eqnarray}}
\newcommand{\bes}{\begin{subequations}}
\newcommand{\ees}{\end{subequations}}

\begin{document}
\interfootnotelinepenalty=10000

\title{Extended Palatini action for general relativity  and the  natural emergence of the 
cosmological constant}

\newcount\hh
\newcount\mm
\mm=\time
\hh=\time
\divide\hh by 60
\divide\mm by 60
\multiply\mm by 60
\mm=-\mm
\advance\mm by \time
\def\hhmm{\number\hh:\ifnum\mm<10{}0\fi\number\mm}

\author{Eran Rosenthal}
\email{eranr@astro.cornell.edu}
\affiliation{Center for Radiophysics and Space Research, Cornell  University, Ithaca, New York, 14853}


\begin{abstract}
In the Palatini action of general relativity 
the connection and the metric are treated as independent dynamical variables. 
Instead of assuming a relation between these quantities, 
the desired relation between them is derived through the Euler-Lagrange equations of the Palatini action.  
In this manuscript we construct an extended Palatini action, where we do not assume any a priori relationship between 
 the connection, the covariant metric tensor, and the contravariant metric tensor. Instead we treat these three quantities  as 
independent dynamical variables. 
We show that this action reproduces the standard Einstein field equations depending on a single metric tensor. 
We further show that in this formulation the cosmological constant has an additional theoretical significance. 
Normally the cosmological constant is added to the Einstein field equations for the purpose of having general 
relativity be consistent with cosmological observations.
In the formulation presented here,  the  
nonvanishing cosmological constant also ensures  the self-consistency of the theory. 
\end{abstract}

\maketitle

\section {Introduction}

In metric theories of gravity,  complete information about the spacetime geometry is
encoded in the components of the metric tensor $g_{\mu\nu}$. 
Yet a complete  field theory on a curved spacetime also requires 
a spacetime connection $\Gamma^{\alpha}_{\ \beta\gamma}$ 
needed for the definition of the covariant derivative, and a contravariant metric tensor $g^{\mu\nu}$ needed 
for the construction of norms  of covariant  vector fields. 
In the   Einstein-Hilbert action of general relativity (GR) it is   {\em assumed} 
that the spacetime connection is the Levi-Civita connection, 
determined by the metric alone, and furthermore it is {\em assumed} that 
the  contravariant  tensor $g^{\mu\nu}$  that appears in the combination  $g^{\mu\nu}R_{\mu\nu}$ in the Einstein-Hilbert action 
  is equal to the inverse matrix of the covariant  metric tensor  ${g}_{\mu\nu}$ also appearing in this action. 
This paper will discuss removing these  implicit assumptions 
and having the relation between the connection $\Gamma^{\alpha}_{\ \beta\gamma}$ and the metric $g_{\mu\nu}$, 
and the relation  between the contravariant metric tensor  $g^{\mu\nu}$ and the covariant metric tensor 
$g_{\mu\nu}$ become  predictions of the theory. 
The first of these assumptions is removed by the Palatini action \cite{palatini}. In this action 
 the connection and the metric are considered to be independent dynamical variables, thereby  producing two sets 
of Euler-Lagrange equations. One set of equations are the Einstein's field equations (depending on the connection and 
the metric), and the other set of equations ensures  that the connection is metric compatible, 
and therefore equals to the Levi-Civita connection. The Palatini formalism has been useful in 
extensions and modification of GR and also in quantization of gravity, see for example Refs. \cite{vollick,flanagan,js,bm,palden}. 
In this paper, we  extend Palatini's formalism and construct an extended Palatini action in which 
 the connection $\Gamma^{\alpha}_{\ \beta\gamma}$, the contravariant 
metric tensor $g^{\mu\nu}$, and the covariant metric tensor $g_{\mu\nu}$ 
that appear in the action are considered to be independent dynamical variables. 
This action may find similar application in modification or quantization of GR.

Our extended Palatini action reproduces GR provided that the  cosmological constant is nonvanishing. 
This is a rather unusual property of this formalism. 
Usually, the cosmological constant is added to the Einstein equations 
in order that the predictions of GR  be consistent with 
the observed accelerated expansion of the Universe, but it does not serve an additional 
theoretical purpose. 
In the formulation presented here, the cosmological constant emerges naturally 
as an ingredient that ensures that the theory is self-consistent, and that the 
Einstein field equations follow from the Euler-Lagrange equations. 
It is interesting to note that  the cosmological constant has  a similar theoretical significance in 
 a very different purely affine gravitational action due to schr\"{o}dinger \cite{sch}.

\section{Action and Field equations}

To see how one arrives to the  gravitational action presented in this paper, it is instructive to 
consider first  the Einstein-Hilbert action. This action  is a functional of the metric alone and it reads
\be\label{eh}
S_{EH}[g_{\mu\nu}]=\frac{1}{16\pi}\int g^{\mu\nu} R_{\mu\nu}(g) \sqrt{-\det(g_{\alpha\beta})} d^4x\,,
\ee
where we set $G=c=1$ throughout.  
Here there are two implicit assumptions. First, it is assumed that  the spacetime connection is the Levi-Civita connection, 
so that the Ricci tensor  $R_{\mu\nu}(g)$  which is normally  determined  by the connection,  
is now  determined by the metric. Second, it is assumed that  the 
contravariant tensor $g^{\mu\nu}$ equals to  the inverse matrix of the metric tensor $g_{\mu\nu}$. 
The Palatini action provides an alternative action,  which reproduces GR 
without making the first implicit assumption. This action reads 
\be\label{palatini}
S_{Palatini} [g_{\mu\nu},\Gamma^{\alpha}_{\ \beta\gamma} ]=
\frac{1}{16\pi}\int g^{\mu\nu} R_{\mu\nu}(\Gamma) \sqrt{-\det(g_{\alpha\beta})} d^4x\,.
\ee
Here the action is a functional of both the metric and the connection\footnote{For completeness we  also need to supplement $S_{EH}$ and 
$S_{Palatini}$ with a matter action $S_M[\psi,g_{\mu\nu}]$, where $\psi$ collectively denotes the matter fields. This introduces some  
ambiguity in the case of the Palatini action, where it is possible to have a matter action  which  
is also a functional of the connection (e.g. in a tetrad formulation for fermions).  Here for simplicity we assume that the matter action depends only 
on the matter fields and the metric.} , 
 and the Ricci tensor  $R_{\mu\nu}(\Gamma)$ by its definition depends only on the 
connection. Assuming that the connection is symmetric $\Gamma^{\alpha}_{\ [\beta\gamma]}=0$ the Euler-Lagrange equations 
  give rise to the Einstein's field equations together with the following equations
\be\label{palatiniel}
\nabla _\epsilon g^{\mu\nu} + g^{\mu\nu}  \left[\partial_{\epsilon}\log\sqrt{-\det(g_{\alpha\beta})}
 -\Gamma^{\alpha}_{\ \alpha\epsilon}\right]=0\,,
\ee
where $\nabla_{\epsilon}$ denotes the standard covariant derivative depending on the connection. 
Relation  (\ref{palatiniel}) imply that connection is metric compatible $\nabla _\epsilon g^{\mu\nu}=0$, and so it  
 ensures that the connection equals to the Levi-Civita connection. 

We now extend the Palatini action and 
 remove the assumption that  the tensor $g^{\mu\nu}$ equals to  the inverse matrix of the metric $g_{\mu\nu}$.
For this purpose we replace the inverse metric tensor  $g^{\mu\nu}$ in the Palatini action (\ref{palatini}) with a new 
independent dynamical variable  denoted $\tilde{g}^{\mu\nu}$. 
To recover  the Einstein field equations we also multiply
the Lagrangian density by $16(g_{\mu\nu} \tilde{g}^{\mu\nu})^{-2}$ and 
introduce a cosmological constant. The resultant  gravitational action reads
\be
S_G[g_{\alpha\beta},\tilde{g}^{\mu\nu},\Gamma^{\rho}_{\ \sigma\eta}]=
\frac{1}{\pi} 
\int \left[\tilde{g}^{\alpha\beta} R_{\alpha\beta}(\Gamma)-2\Lambda\right]
(g_{\mu\nu} \tilde{g}^{\mu\nu})^{-2} \sqrt{-\det(g_{\rho\sigma})}d^4x \,.
\ee
Here $g_{\alpha\beta}$, $\tilde{g}^{\mu\nu}$ and $\Gamma^{\alpha}_{\ \beta\gamma}$ are independent dynamical variables and 
 we assume  that the connection is symmetric in its lower indices.
Notice that from its definition  this action has two independent 
metric tensors:  a contravariant tensor $\tilde{g}^{\mu\nu}$ and a covariant tensor 
  ${g}_{\mu\nu}$.  For clarity we shall not raise or lower indices with any 
of these metric tensors.  
In addition we introduce the following matter action 
\be\label{sm}
S_M=S_M[\tilde{g}^{\mu\nu},\psi]\,.
\ee
Here $\psi$ collectively denote the matter fields.  Notice that this action does not 
depend on the metric $g_{\mu\nu}$.
The complete action is given by 
\be
S=S_G+S_M\,.
\ee
 
The action $S$ gives rise to the following  Euler-Lagrange equations 
\begin{eqnarray}\label{el1}
&&(\tilde{g}^{\alpha\beta} R_{\alpha\beta}-2\Lambda) 
(g^{\mu\nu}-\frac{4}{d} \tilde{g}^{\mu\nu} )=0\,,\\ \label{el2}
&&R_{\alpha\beta}- 
\frac{2}{d} g_{\alpha\beta}(\tilde{g}^{\mu\nu}R_{\mu\nu} -2\Lambda)  
=\frac{\pi d^2}{2}(  \tilde{g} /g )^{1/2} T_{\alpha\beta}\,,\\ \label{el3}
&&\nabla _\epsilon \tilde{g}^{\alpha\beta} + \tilde{g}^{\alpha\beta}  
\left[\partial_{\epsilon} \log (d^{-2}\sqrt{- g})   -\Gamma^{\gamma}_{\ \gamma\epsilon}\right]=0\,,\\ \label{el4}
&&\frac {\delta S_M}{\delta \psi}=0\,.
\end{eqnarray}
Here  we have introduced the tensors $g^{\alpha\beta}$ and $\tilde{g}_{\mu\nu}$ which are defined 
by 
$g^{\gamma\alpha}g_{\alpha\beta}=\delta^{\gamma}_{\beta}$ and $\tilde{g}^{\gamma\alpha}\tilde{g}_{\alpha\beta}=\delta^{\gamma}_{\beta}$, respectively. 
For brevity we have also introduced the notation
$d\equiv g_{\rho\sigma} \tilde{g}^{\rho\sigma}$, 
$\tilde{g}\equiv \det(\tilde{g}_{\alpha\beta})$, and  $g\equiv\det (g_{\alpha\beta})$. 
In addition, we have defined the energy-momentum tensor $T_{\alpha\beta}$ trough 
\[
\delta S_M=-\frac{1}{2}\int T_{\alpha\beta} \delta \tilde{g}^{\alpha\beta} \sqrt{-\tilde{g}} d^4x \,.
\]

Contracting  Eq. (\ref{el2}) with $\tilde{g}^{\alpha\beta}$ gives
\[
R_{\alpha\beta}\tilde{g}^{\alpha\beta}=-\frac{\pi}{2}  d^2
(\tilde{g}  /  {g})^{1/2} T_{\mu\nu}\tilde{g}^{\mu\nu}
+4\Lambda\,.
\]
Substituting this equation into Eq. (\ref{el1}) yields 
\be\label{metricseq}
[2\Lambda-\frac{\pi}{2}d^2(\tilde{g}  /  {g})^{1/2} T_{\alpha\beta}\tilde{g}^{\alpha\beta}](g^{\mu\nu}-\frac{4}{d}  \tilde{g}^{\mu\nu} )=0\,.
\ee
Notice that a contraction of Eq. (\ref{metricseq}) with the metric ${g}_{\mu\nu}$  yields
 an identity, thus providing a consistency check for this equation. To understand the consequences of 
Eq. (\ref{metricseq}) let us consider first an empty spacetime region with a nonvanishing cosmological constant,  
 so that $T_{\alpha\beta}=0$. In this region Eq. (\ref{metricseq}) is reduced to
\be\label{metrics2}
\Lambda(g^{\mu\nu}-\frac{4}{d}  \tilde{g}^{\mu\nu} )=0\,.
\ee
Eq. (\ref{metrics2}) provides a relation between the metric $g_{\alpha\beta}$ 
and the metric $\tilde{g}^{\alpha\beta}$ provided that the cosmological constant does not vanish. 
Notice that if the cosmological constant does vanish 
the remaining Euler-Lagrange equations (\ref{el2},\ref{el3},\ref{el4}) 
are insufficient to determine  any of the metric tensors. 
We now demand that theory be sufficiently general 
and admit solutions with empty regions of spacetime. By virtue of 
the above analysis  this requirement forces us to impose a nonvanishing cosmological constant $\Lambda\ne0$.
Next, let us focus our attention to a region with matter, where $T_{\alpha\beta}\ne 0$.
Here there are  three possibilities: First, the expression in the square brackets in Eq. (\ref{metricseq}) may 
be different from zero. In this case Eq. (\ref{metrics2}) remains valid. Second, 
the expression in the square brackets may be different from zero everywhere 
except for a submanifold with dimensionality smaller then 
four (e.g. this expression may vanish on a three dimensional hypersurface). 
In this case, by continuity, Eq. (\ref{metrics2}) remains valid. 
Third, the expression in the square brackets may vanish in a four dimensional submanifold. 
In this case Eq. (\ref{metrics2}) does not 
follow from Eq. (\ref{metricseq}). However, this case is nongeneric and demands  
 fine tuning, and so it does not correspond 
to a realistic physical configuration. Furthermore, given such a fine tuned configuration, we can 
normally introduce an infinitesimal (and so experimentally unobserved) modification to $T_{\alpha\beta}$ 
such that the expression in the square brackets would no longer vanish. For these reasons we  
discard these pathological field configurations.
We now proceed to derive the Einstein field equations.

It is useful to rewrite (\ref{metricseq}) as 
\be\label{metricseq2}
\tilde{g}_{\mu\nu}=\frac{4}{ d}g_{\mu\nu}\,.
\ee
Substituting  Eq. (\ref{metricseq2}) into Eq. (\ref{el3}) gives 
\be\label{metriccomp}
\nabla _\epsilon \tilde{g}^{\alpha\beta} + \tilde{g}^{\alpha\beta}  
\left[\partial_{\epsilon}\log\sqrt{-\tilde{g}}  -\Gamma^{\gamma}_{\ \gamma\epsilon}\right]=0\,.
\ee
Eq. (\ref{metriccomp}) has the same form as Eq. (\ref{palatiniel}) produced by 
 the standard Palatini action (\ref{palatini}) and so it implies that the connection equals 
to the Levi-Civita connection evaluated with the metric $\tilde{g}^{\mu\nu}$. 
  Substituting  Eq. (\ref{metricseq2}) into Eq. (\ref{el2}) yields the standard Einstein field equations
\be\label{efe}
R_{\alpha\beta}(\tilde{g})-\frac{1}{2} \tilde{g}_{\alpha\beta} \tilde{g}^{\mu\nu}R_{\mu\nu}(\tilde{g})  +\tilde{g}_{\alpha\beta}\Lambda  =
8\pi T_{\alpha \beta}\,.
\ee
Here $R_{\alpha\beta}(\tilde{g})$ denotes the Ricci tensor evaluated from
the metric $\tilde{g}_{\alpha\beta}$. We see that  Eq. (\ref{efe}) depends on  the metric $\tilde{g}_{\mu\nu}$ and is independent of 
the metric   ${g}_{\mu\nu}$. 
In other words {\em we have found that the contravariant metric tensor  $\tilde{g}^{\mu\nu}$  that appears 
in the action $S$ equals to the inverse matrix of the metric  $\tilde{g}_{\mu\nu}$ that satisfies 
the  Einstein field equations}  (\ref{efe}).  
Notice that the  Euler-Lagrange equations (\ref{el1},\ref{el2},\ref{el3},\ref{el4}) have been reduced to 
 three differential equations (\ref{el4},\ref{metriccomp} ,\ref{efe}) and one algebraic relation (\ref{metricseq2}). 
The fact that these differential equations  depend only on  
the variables  $\tilde{g}^{\mu\nu}$ and $\psi$ but are independent of  the variable  ${g}_{\mu\nu}$,  signals
that all of the independent physical degrees of freedom in this theory are encoded in the fields $\tilde{g}^{\mu\nu}$ and $\psi$.   
Furthermore, the absence of the metric ${g}_{\mu\nu}$ from the field equation that depend on the 
matter fields,  Eqs. (\ref{el4},\ref{efe}), means that the 
metric ${g}_{\mu\nu}$  can not have  any observational consequences. 
To further clarify this point notice that  using 
 Eq. (\ref{metricseq2}) we can express 
the metric $g_{\mu\nu}$ as $g_{\mu\nu}=\Omega^2(x) \tilde{g}_{\mu\nu}$, where $\Omega^2(x)$ is a conformal factor.
However, the conformal factor  $\Omega^2(x)$  can not be determined by the Euler-Lagrange equations   
(\ref{el1}-\ref{el4}).
To see this, notice that 
the action $S$  is invariant under the  conformal transformation $g_{\mu\nu}\rightarrow \Omega^2(x) {g}_{\mu\nu}$, where 
$\psi$ and $\tilde{g}^{\mu\nu}$  are kept fixed; implying  that  
this conformal transformation is a symmetry of the theory. 
This ambiguity in the determination of  the metric  ${g}_{\mu\nu}$, as opposed to the well posed initial value formulation for 
the metric $\tilde{g}_{\mu\nu}$  reinforces the interpretation that  $\tilde{g}_{\mu\nu}$ is  the only physical metric in this theory.

\section {Conclusions}

By extending the Palatini action, 
we have shown that the Einstein field equations 
can be derived from a gravitational action
that does not make  any prior 
 assumptions about the relation between the contravariant metric  and  
the covariant metric.  
In this theory the Einstein field equations (\ref{efe})  are valid only if the 
cosmological constant does not vanish.  

\acknowledgments 
I thank E. Flanagan for discussions and for comments on an earlier version of this manuscript. 
This work was supported by NSF  grants, PHY 0457200 and PHY 0757735.

\newcommand{\apjl}{Astrophys. J. Lett.}
\newcommand{\aap}{Astron. and Astrophys.}
\newcommand{\cmp}{Commun. Math. Phys.}
\newcommand{\grg}{Gen. Rel. Grav.}
\newcommand{\lr}{Living Reviews in Relativity}
\newcommand{\mnras}{Mon. Not. Roy. Astr. Soc.}
\newcommand{\pr}{Phys. Rev.}
\newcommand{\prsl}{Proc. R. Soc. Lond. A}
\newcommand{\ptrsl}{Phil. Trans. Roy. Soc. London}

\end{document}